\documentclass[%
 reprint,nofootinbib,
 amsmath,amssymb,
 aps,twocolumn,pra
]{revtex4-2}

\usepackage{graphicx}
\usepackage{hyperref}
\usepackage{multirow}
\usepackage{color}
\usepackage{amsfonts}

\newcommand{\eqn}[1]{\begin{eqnarray} #1 \end{eqnarray}}
\newcommand{\tit}[1]{\textit{#1}}
\newcommand{\tbf}[1]{\textbf{#1}}
\newcommand{\trm}[1]{\textrm{#1}}

\definecolor{airforceblue}{rgb}{0.36, 0.54, 0.66}

\begin{document}

\title{QBism and Relational Quantum Mechanics compared}

\author{Jacques Pienaar}
\affiliation{
QBism group, University of Massachusetts Boston, MA, USA.
}


\date{\today}



\begin{abstract}
The subjective Bayesian interpretation of quantum mechanics (QBism) and Rovelli's relational interpretation of quantum mechanics (RQM) are both notable for embracing the radical idea that measurement outcomes correspond to events whose occurrence (or not) is relative to an observer. Here we provide a detailed study of their similarities and especially their differences.
\end{abstract}

\maketitle

\section{Introduction}

In this paper we compare the subjective Bayesian interpretation of quantum mechanics (QBism) \cite{ QB_PERIMETER,QB_FUCHS_2011,QB_FUCHS_SCHACK,QB_FDR2014,QB_Fuchs_PR,QB_FuchsNWB,QBISM_HERO,QB_BCS_EnRoute,QB_FAQBism,QB_VITALITY,QB_Pienaar2020,QB_FELLOW,QB_COHERENCE} with Rovelli's relational interpretation of quantum mechanics (RQM) \cite{ROVELLI_96,RQM_LAUDISA_EPR,RQM_EPR,RQM_BROWN,RQM_VANF,RQM_DORATO,RQM_RUYANT,RQM_TRASSINELLI,RQM_KRISMER,RQM_LAUDISA_OPEN,RQM_LOCALITY,RQM_PIENAAR_COMMENT,RQM_YANG,SEP_RQM,RQM_BIRDS,RQM_CALOSI,RQM_BUNDLE,RQM_STABLE_FACTS}. Both interpretations are notable for endorsing the radical principle of \tit{relative facts}, which holds that the question whether or not a physical quantity has been observed to take a definite value only has an answer relative to a specified observer. Although QBism and RQM are not alone in their endorsement of relative facts (see for instance Refs.\ \cite{BW_BRUKNER_NATURE, BW_BRUKNER, BRUKNER_17, MUELLER_RELFACTS}), they are distinctive for being the longest-running interpretations that take relative facts as a core principle.

The present work highlights key similarities and especially differences between QBism and RQM, with the goal of helping readers already somewhat sympathetic to both interpretations to make an informed judgement as to which one best fits their own principles. We therefore adopt a non-critical stance towards both interpretations; a companion paper, Ref. \cite{PIENAAR_QUINTET}, is devoted to a critical analysis of RQM. We therefore also assume the reader has some familiarity with the literature on QBism and RQM; for a review of the essentials we recommend Refs.\ \cite{QBISM_HERO,QB_FAQBism} and Refs.\ \cite{RQM_BIRDS, SEP_RQM}, respectively. 

The viability of interpretations that endorse relative facts has come to the fore in recent debates surrounding the famous Wigner's friend thought experiment, which has been revived in a number of surprising new variations \cite{BW_BONG,BW_BRUKNER,QB_FELLOW,BW_CAVALCANTI,BW_COMMENT,BW_BRUK_BAUM,BW_FR,BW_HEALEY,BW_PROIETTI,BW_BAUMANN_WOLF}. The central issue in both classic and modern variants of the thought experiment is that a measurement ostensibly carried to completion by Wigner's friend can in principle be fully reversed by Wigner, calling into question the factual nature of the friend's measurement outcome. In these debates it has become clear that it is possible to escape both the traditional and recent versions of the dilemma by maintaining that the friend's measurement outcome can be a fact relative to the friend even when it is not a fact relative to Wigner, hence endorsing relative facts.

QBism and RQM belong to a group of interpretations that trace their lineage to the `Copenhagen interpretation' as promoted (not always consistently or in unison) by Bohr, Heisenberg and Pauli. The hints of disagreements between members of the original Copenhagen school have since been further magnified over time, as different authors following in the footsteps of Copenhagenism have sought to address the modern no-go theorems in increasingly divergent ways. We are left today with a motley band of what Matt Leifer calls ``Copenhagenish" interpretations \cite{LEIFER_WIGNER}, which bear only a family resemblance to one another and to the original Copenhagen school. The most prominent of them besides QBism and RQM include Healey's pragmatist interpretation \cite{BW_HEALEY}; the `neo-Copenhagenism' of Brukner and Zeilinger \cite{BRUKZEIL_03, BRUKNER_17}; and the `information theoretic' interpretation of Bub and Pitowski \cite{BUB_PITOWSKI, BUB_19}.

Copenhagenish interpretations are loosely characterized by their adherence to the following four principles (stated in more detail in Sec.\ \ref{sec:fourprinciples}): 
\begin{enumerate}
\item Measurement outcomes for a given observer are \tit{unique}, i.e.\ contrary to many-worlds interpretations, there do not exist multiple copies of the same observer that observe different outcomes; 
\item The quantum state is of a \tit{broadly epistemic} character, i.e.\ it represents `information', `knowledge', or `beliefs'; 
\item Quantum theory is a universally applicable theory, i.e.\ it can be consistently applied to arbitrary scales, systems, and parameter regimes; 
\item Quantum theory is a complete theory, i.e.\ does not require completion by the addition of supplemental or `hidden' variables.
\end{enumerate}

Note that since both QBism and RQM adhere to the principles (i)-(iv) shared by all Copenhagenish theories, this is already enough to explain most of their notable similarities. In addition, both interpretations endorse relative facts, which unites them against almost all other Copenhagenish interpretations\footnote{A notable exception being Brukner's interpretation \cite{BRUKNER_17,BW_BRUKNER}.}. However, the apparent alliance is deceptive: QBism and RQM give very different accounts of the meaning of quantum states and measurement outcomes, leading them to endorse fundamentally different understandings of reality. It is these fundamental differences between QBism and RQM that will be the main focus of the present work.

Our task of comparing QBism and RQM is complicated by the fact that both interpretations have evolved over time, and have benefited (and suffered) from contributions made by other authors who did not entirely share the same vision as the originators of the interpretations. When conflicts or doubts arise, we shall give precedence to the authority of papers co-authored by the originators of each interpretation (Chris Fuchs and R\"{u}diger Schack for QBism, and Carlo Rovelli for RQM), and among those we shall give precedence to more recent papers.

The paper is structured in three parts. In Sec.\ \ref{sec:fourprinciples} we compare QBism and RQM according to the different ways in which they conform to Leifer's four principles (i)-(iv) of Copenhagenish interpretations. In Sec.\ \ref{sec:lawsnorms} we point out QBism's and RQM's different attitudes towards quantum `laws'. Finally, in Sec.\ \ref{sec:agreement} we discuss a key difference in how they treat the possibility of agreement between observers.

\section{A difference in four principles \label{sec:fourprinciples}}

We begin with the first principle of Copenhagenish interpretations:

\begin{quote}
\tbf{(i)}\ Measurement outcomes for a given observer are \tit{unique}. There only exists one instance of any given observer, any measurement they perform has exactly one outcome relative to them.
\end{quote}

\tit{RQM:} The uniqueness of measurement outcomes in RQM derives from the uniqueness of the `relative facts' produced in \tit{interaction events}. An interaction event takes place when two systems interact and thereby exchange information, i.e.\ become correlated in certain degrees of freedom. When this happens, RQM stipulates that we can take one of the systems to be an `observer' and the other to be an `observed system', which entails that we can assign a set of \tit{unique} values to the relevant degrees of freedom. These unique values then represent \tit{relative facts} about the observed system, relative to the \tit{context} of the observing system. Di Biagio and Rovelli put it concisely in Ref. \cite{RQM_STABLE_FACTS}: 
\begin{quote}
Relative facts are defined to happen whenever a physical system interacts with another physical system. Consider two systems $S$ and $F$. If an interaction affects $F$ in a manner that depends on the value of a certain variable $L_S$ of $S$, then the value of $L_S$ is a fact relative to $F$. This is true by definition irrespectively of whether $F$ is a classical system or not. That is, whenever the two systems interact, the value of the variable $L_S$ becomes a fact \tit{relative} to $F$.
\end{quote}  

Thus, in RQM, the uniqueness of measurement outcomes is a consequence of the fundamental process of the \tit{becoming of relative facts}, which is defined as a process that, in the fundamental interaction event, spontaneously produces \tit{unique values} of the observed variables relative to the context of the observing system. 

\tit{QBism:} In QBism, the notion of \tit{measurement} is taken as fundamental. A measurement is more than just a physical interaction: it involves an \tit{action} by the observer (their choice of which measurement to enact) and a corresponding \tit{result} (the measurement outcome). For this reason, QBism prefers to speak of an \tit{agent} rather than an `observer', to emphasize that the agent makes choices and takes voluntary actions. (We will continue to use the word `observer' as a neutral term that encompasses both the observers of RQM and the agents of QBism, and will use `agent' only when making statements specific to QBism).

In QBism, a measurement outcome is interpreted as an agent's subjective experience of something that happens to them as a result of some action they took. Uniqueness for the QBist follows from reflecting upon the nature of experience itself: when I have an experience, I only experience one out of the many possible experiences that I could have experienced. Moreover, the details of the experience allow me to identify myself as the unique `self' capable of having that precise experience\footnote{In QBism I cannot meaningfully posit that there is another version of `me' who had a different experience; per definition that other person would not be `me'.}.

\tit{Remarks:} In both QBism and RQM, we see that measurement outcomes are treated as fundamental events, in which certain variables of the observed system spontaneously acquire unique values relative to the observer. However, beyond this similarity, the accounts could not be more different. In RQM, these outcomes are impersonal naturalistic events: they simply occur between any pair of interacting systems, either one of which could serve as the `observer' of the other one. By contrast, a measurement outcome in QBism is an experience, hence requires an agent that is capable of having experiences; thus to the extent that we don't consider rocks and electrons to have experiences, they are not agents in QBism. 

Furthermore, in RQM the variables that can be `measured' by the observer are determined to include all those degrees of freedom of the system to which the observer is sensitive, i.e.\ to which the observer's degrees of freedom are correlated through the interaction. By contrast, in QBism the measured variables are the ones which the observer is capable of \tit{perceiving}. Thus, although my body is currently interacting with the ambient background radiation, a QBist would not say I am measuring it unless I am aware of it in some way, eg. if I happen to be holding a Geiger counter whose audible clicks I can discern.

\begin{quote}
\tbf{(ii)}\ The quantum state represents `information'. That is, the quantum state is not itself a physically existent \tit{thing} in the world, but has an epistemic character in that it represents \tit{information about} things or propositions that might obtain in the world. 
\end{quote}

A clarification is in order: to say that the quantum state represents information of an `epistemic' character does not commit us to a so-called \tit{psi-epistemic} interpretation, according to which the quantum state represents partial information about some hypothetically pre-existent hidden variables~\cite{HARSPEK}; we refer to such views epistemic in a narrow sense. Since Copenhagenish interpretations in general also subscribe to principle (iv), thereby rejecting hidden variables, they must necessarily reject this narrow interpretation of `epistemic'. But \tit{broadly} epistemic interpretations remain viable, in which the information encapsulated in the quantum state is about something other than hidden variables.

Bell famously challenged the evasive meaning of `information' by asking: ``\tit{whose} information? Information about \tit{what}?" \cite{BELL_AM}. Provided that we remain deliberately vague about the meaning of `observer', and `probability', both QBism and RQM endorse the same general answer to Bell: 

\begin{quote}
The quantum state represents a convenient `book-keeping device' which encapsulates information about an \tit{observer's probability assignments} for the \tit{outcomes of possible future measurements} that the same observer could perform on the system.
\end{quote}

This information is `epistemic' in the sense that it depends on the particular observer's perspective and history, as regards to the system in question. Moreover, once the relevant probabilities are defined, both QBism and RQM may take advantage of the full arsenal of tools from information theory following Shannon. For example, we can use the probabilities to define entropies and conditional entropies of the various system variables, and then use these to define quantities such as self-information, mutual information, and so on. The key difference between QBism and RQM here lies only in their different interpretations of the terms `observer' and `probability'. We now discuss these differences.

\tit{RQM:} Since observers are nothing more than physical systems in RQM, the information that one system has about another is understood to refer to correlations between variables representing physical properties of the two systems:

\begin{quote}
[W]e say that a variable $O_A$ of a system has information about a variable $A$ of another system if the values that $A$ and $O_A$ can take are correlated. In the spirit of Shannon, this is a very weak definition of information that has no mentalistic, semantic, or cognitive aspects. --- \cite{SEP_RQM}.
\end{quote}

Note that, notwithstanding the ``spirit of Shannon", the lack of mentalistic, semantic, or cognitive aspects does not stem from Shannon's definition of information, which is compatible with any interpretation of probability; rather it stems from RQM's assumption that the probabilities represent \tit{objective uncertainty} about physical systems, implying that they have no mentalistic, semantic, or cognitive aspects.

We have mentioned the fact that the quantum state assigned to a system depends somehow on the `perspective' of the observer. Since the observer is itself a physical system in RQM, this `perspective' must be understood in terms of objective physical facts associated with the observing system, i.e.\ without requiring that it have anything like a mind or the ability to have experiences. The solution in RQM is to define the observer's `perspective' as the set of unique values assigned to all of its previous interactions with all relevant systems (i.e.\ the outcomes of past measurements on them) \cite{SEP_RQM}. In theory, this compendium of `observer-relative facts' is sufficient to uniquely specify the quantum state that the observer must assign to any given system. As Laudisa and Rovelli put it:

\begin{quote}
[The quantum state is] a mathematical device that refers to two systems, not a single one. It codes the values of the variables of the first that have been actualised in interacting with the second [\dots] [It] therefore codes anything we can predict regarding the future values of these variables, relative to the second system. The state $\Psi$, in other words, can be interpreted as nothing more than a compendium of information assumed, known, or gathered through measurements, determined entirely by a specific history of interactions: the interactions between the system and a second ‘observing’ system. --- \cite{SEP_RQM}.
\end{quote}

This account suggests that the quantum state relative to an observer's perspective in RQM may be determined along the following lines: one begins with a conventional state assigned to the relevant system that applies when the observer has not yet had any interactions that are informative about the system (RQM does not specify what this state should be, but for finite-dimensional systems the maximally mixed density operator seems an obvious choice). One then employs the L\"{u}ders rule to update the state in light of subsequent interaction events in the observer's perspective, taking them in chronological order\footnote{Rovelli has implied in Ref.\cite{RQM_BIRDS} and elsewhere that RQM is agnostic to the direction of time. This implies that we may choose the observer's proper time to increase in either direction, corresponding to different ways of defining the observer's `history' and `future', hence different ways of defining their perspective and state assignments.}. Finally, the state of the system relative to the observer is defined as the state reached after conditioning on the observer's perspective. 

The important point is that there is only one objectively correct quantum state assignment from a given observer's perspective, and subjective elements play no role in determining it.

\tit{QBism:} For the QBist, as we have seen, the observer for whom a quantum state is defined cannot be just any physical system, for it must at least be capable of having experiences. In fact QBism demands even more: the observer must qualify as an \tit{agent} in the decision theoretic sense, which means they must be capable of reasoning using probabilites, and must strive to make \tit{coherent} decisions. Moreover, within this decision-theoretic framework, probabilities (including those that are encapsulated by the quantum state) have a \tit{subjective Bayesian} interpretation: they represent an agent's quantified \tit{subjective degrees of belief} about the likelihood of the outcomes of their possible future measurements.

In QBism, the quantum state that an agent assigns to a system can be determined by a process superficially similar to that of RQM: an initial \tit{prior} state assignment is updated in light of the agent's past record of measurement outcomes. However, in QBism the choice of appropriate \tit{prior} is not given by quantum theory -- rather it is supplied by the agent's own initial beliefs, which may include such `subjective' considerations as the agent's general worldview, temperament, etc, and can therefore differ wildly even between two agents who possess the same initial data about the system. Moreover, the rule for updating the state depends on the agent's beliefs about how the measuring apparatus functions, and is represented in general by an arbitrary quantum map. It follows that in QBism the agent's final quantum state assignment is entirely a function of their subjective beliefs; there is not one `objectively correct' state assignment.

\tit{Remarks:} Someone having only a superficial acquaintance with QBism and RQM might have the impression that the \tit{only} essential difference between them is their interpretation of probability. However, our preceding analysis shows why the rift between the two interpretations is much more significant than that. 

The interpretations of probability favoured by QBism and RQM are not incidental features of each, but are consequences of deeper considerations of a distinctly ontological nature. In RQM, measurement outcomes are interaction events that retain the impersonal, materialistic connotations of `events' in textbook physics, that is, spontaneous natural occurrences (albeit relational ones) for which any reference to consciousness, perception, experience, or other subjective qualities has been eliminated. It follows that although probabilities in RQM are not themselves ontological objects, they do possess a certain kind of ontological significance, in that they can still be meaningfully defined in situations where there are no agents around, such as the case of one atom observing another. This implies that RQM cannot dispense with an objective interpretation of probability that governs inanimate matter.

It follows that although subjective probabilities can be included within RQM to represent the beliefs of any rational agents who might be around, these subjective probabilities must co-exist together with the more fundamental objective probabilities that reflect the true likelihood of natural events as prescribed by quantum theory.

By contrast, QBism's assertion that measurement outcomes are experiences of an agent lead it to a quite different conclusion, namely, that probabilities (hence quantum states) do not have \tit{any} ontological significance. Probabilities (and by extension quantum states) are necessarily subjective: they are the products of agents' special ability to reflect upon, reason about, and anticipate their experiences. It follows that in the absence of any beings capable of such reflection, there can be no probabilities at all. 

Notice, however, that QBism treats probabilities (which are \tit{about} possible experiences) differently from the experiences themselves. Just because probabilities have no ontological content in QBism does not mean that experiences themselves have no ontological content. In fact, Fuchs has often suggested that the concept of `measurement outcomes' can be extended beyond agents' experiences to encompass a more general notion of `little moments of creation' \cite{QBISM_HERO, QB_FuchsNWB, QB_Fuchs_PR} that can occur whether or not any agents are present. Thus QBism holds the door open to the possibility of an `ontological substrate' that might exist without agents or probabilities.

What is the nature of this ontological substrate? Although the answer is an active topic of debate among QBists and philosophers, one thing is certain: the ontological substrate of QBism cannot be of a purely `mechanical' or `physical' nature. The reason is simple: since QBism takes measurement outcomes to be experiences of agents, what QBism's ontology seeks to explain is not just how `physical phenomena' arise, but rather, \tit{how physical phenomena co-arise together with agents' experiences of them}. As Fuchs explains: 

\begin{quote}
    The case QBism makes before the forum is this. What the quantum agent -- the protagonist in the drama of any application of quantum theory -- is ultimately doing is hitching a ride with a \tit{new kind of ontology or metaphysic}. [\dots] This suggests that the ontological abstracta of our decision-theoretic conception of quantum theory is neither the agent, nor the object, but something that can be polarized in one direction or the other depending upon the attending analysis one gives it. William James’s `pure experience' and Alfred North Whitehead’s `actual occasions' have something of this character [\dots]. -- \cite{QB_FuchsNWB}, p32.
\end{quote}

QBism therefore requires a `new kind of ontology' capable of giving rise to both agency and to the objects of agency, treating them as equally fundamental. Such a view of ontology is not alien to philosophy, as is evidenced by QBism's alliance with the philosophies of Dewey, James, Whitehead, and more recently Husserl and Merleau-Ponty; but it is a radical departure from the strictly physicalist ontology endorsed by nearly all other quantum interpretations, including RQM.

The above considerations show that the divide between QBism and RQM goes much deeper than their different preferences for interpretations of probability. One could certainly imagine reformulating QBism so as to make agents' probability assignments objective, but this would not be enough to make QBism resemble RQM even slightly\footnote{More plausibly, doing so might result in something resembling Healey's pragmatist interpretation.}: probabilities would still be meaningless without agents, and making them objective could not bring QBism's `moments of creation' any closer to RQM's `interaction events'.

\begin{quote}
\tbf{(iii)}\ Quantum theory is a universally applicable theory, i.e.\ it can be consistently applied to arbitrary scales, systems, and parameter regimes.
\end{quote}

\tit{RQM:} In RQM quantum theory as a whole is regarded as capturing statements about the nature of reality. It is, in this sense, descriptive of the material universe, relative to any given observer within that universe. One cannot, of course, contemplate an observer outside the universe, because observers are themselves physical systems within the universe. Thus, one cannot meaningfully assign a quantum state to the universe as a whole. Nevertheless, the theory is universal in that it allows any \tit{part} of the universe to play the role of an observer of \tit{any other part}. The universality of the theory expresses this fact: the theory takes the same form, regardless of the particulars of the physical systems involved. Although it does not describe the universe as seen from nowhere, RQM purports to describe all observer-system relationships that exist in the universe, and moreover it declares that these relations are all that exist at the fundamental level. That is what the claim of universality means in RQM.

\tit{QBism:} In QBism quantum theory is not a descriptive theory at all -- it is a decision theory. A decision theory is like a guide to acting in the world: it neither describes the world directly, nor does it directly describe agents who are the users of the theory. A decision theory is a set of normative rules, which may be thought of as rational \tit{tools}, that are employed by agents in order to make \tit{coherent} judgements and actions, where for present purposes we may think of `coherent' as meaning both rational and internally consistent. Being a \tit{tool} of this kind, quantum theory is no more a description of reality than a hand-saw is a description of wood. 

To be sure, by analyzing the blade of a hand-saw, one might hope to infer something about the material that it is used to cut. Moreover, quantum theory is not a description of the \tit{agent} any more than the hand-saw is a description of a hand, but again, we might hope to \tit{infer} something about the hand by examining the handle of the hand-saw. QBism's methodological project is to better understand both agents and reality by carefully examining the structure of quantum theory as a tool that agents use to make good decisions in their encounters with reality. 

What, then, does the claim of universality mean for the QBist? It simply means that quantum theory is a tool that is \tit{universally applicable}, in two senses: first, \tit{any} agent can use it; and second, it can be used upon \tit{any} part of reality (including other agents).

\tit{Remarks:} Although the meaning of universality is different in both QBism and RQM, they tend to agree on its consequences, and on the responses to certain criticisms that are sometimes made in light of these consequences. 
One consequence is that both QBism and RQM cannot meaningfully assign a quantum state to the whole universe, which has led some critics to ask how cosmology can be possible. But nothing prevents one from meaningfully discussing \tit{models} of the whole universe, provided that we interpret these models as ultimately referring to and drawing their justification from observations that we make as observers within the universe \cite{QB_PERIMETER, QBISM_HERO}.

Another consequence is that both QBism and RQM do not allow an observer to assign a quantum state to themselves. This is because both interpretations require the specification of an observer/system split in order to make meaningful statements about physics. In RQM the issue is a matter of deciding which piece of matter is to serve in the role of the `frame' or `context' and which piece of matter is to be described as a system relative to that context. In QBism the issue is that the usage of quantum theory -- like the usage of any tool -- requires both a user and a system upon which the tool is to be used. Either way, a system cannot serve simultaneously in both roles, and when it serves in the role of context or user, it is by definition excluded from the quantum state.

Contrary to some critics' impressions, the existence in QBism and RQM of an observer/system split is not an obstacle to universality in either interpretation, because both of them remain flexible about where to draw the boundary between observer and observed system. An observer is free to take any part of themselves as a separate object, upon which they can perform measurements -- this only requires regarding the relevant part as external to themselves for the purposes of the analysis. Indeed, self-identity of an observer is something ambiguous and fluid in both interpretations, though for different reasons. In RQM, the physical boundary that encloses the matter which makes up an observer can be redefined; in QBism, the sensory boundary that defines the limits of the agent's perceptions (i.e.\ their body) can similarly be redefined \cite{QB_Pienaar2020}. 

It has sometimes been argued that the Wigner's friend thought experiment forces the friend to assign a quantum state to herself, in order to make the `correct' predictions \cite{BW_BRUK_BAUM}. Both QBism and RQM reject this claim, for essentially the same reason: it presumes that the correctness of a probability assignment can be established independently of the differing perspectives of the two observers in the experiment. But since the friend and Wigner have different perspectives, QBism and RQM allow them to make different probability assignments without either of them having to be `incorrect' \cite{QB_FELLOW, RQM_STABLE_FACTS}.

Finally, Cavalcanti has recently shown that in QBism the friend's supposed incorrectness cannot actually lead them into negative consequences, say, by losing money on bets which they might make based on their probability assignments~\cite{BW_CAVALCANTI}. One might expect a parallel argument to hold in RQM, establishing that both Wigner and the friend's state assignments are the best possible assignments given the information in principle available to each of them.

\begin{quote}
\tbf{(iv)}\ Quantum theory is a complete theory, i.e.\ does not require completion by the addition of supplemental or `hidden' variables.
\end{quote}

The rejection of hidden variables implies that the outcome of any given measurement cannot be interpreted as revealing the value of some unobserved pre-existing element of reality. Both QBism and RQM opt for the simplest version of this doctrine, which outright denies the existence of unobserved elements of reality. Thus, both would decline to posit the existence of anything `in between' measurement outcomes, so to speak. 

This still leaves open the question of how to interpret the fact (implied by the quantum formalism) that one cannot be absolutely certain about the outcomes for \tit{every} possible measurement on a system. Since hidden variables are excluded, this cannot be attributed to mere ignorance of hypothesized ontological facts: rather, it must in some way be taken to indicate the \tit{absence} of such facts at the ontological level. QBism and RQM understand this `essential absence' in subtly different ways.

\tit{RQM:} In the earliest paper on RQM, Rovelli postulates: ``it is always possible to acquire \tit{new} information about a system", which he emphasizes is a ``fully experimental" postulate. The suggestion is that no matter how many measurements an observer performs on a system, they must always assign it a quantum state that implies uncertainty about the values of some future possible measurements. Moreover, since the amount of information associated to a system is postulated to be finite in quantum theory, this implies that the new information that comes into being in a physical interaction is \tit{genuinely new}, that is, cannot be accounted for by the set of all previously existent facts about the system. 

This \tit{genuine novelty} has an `objective' character in RQM, since it pertains to the objective relation between an observing and an observed system, and is definable in the absence of any `agents'. Thus RQM cannot dispense with the need for objective probabilities, which may be thought of as referring to the fundamental indeterminacy of quantum relations. Dorato has suggested that it be thought of as a \tit{dispositional} quality (i.e.\ endorsing a objective dispositionalist interpretation of probability), and Laudisa and Rovelli agree that this ``fits particularly well in the context of RQM"\cite{SEP_RQM}. However, other objective definitions of probability may be considered. 

One notable consequence of taking a fundamentally objective view of probabilities is that, in the special case where an observer assigns an objective probability 1 to the outcome of a particular measurement they could perform, then that outcome \tit{must} actually occur if the observer subsequently performs the measurement. Although this is reminiscent of the so-called eigenstate-eigenvalue link, note that in keeping with principle (iv) of Copenhagenish interpretations (no hidden variables), the probability-1 assignment cannot imply the existence of an element of reality prior to (or after) the interaction event; it only serves to determine the element of reality in the very moment of interaction. The eigenstate-eigenvalue link therefore retains its significance in RQM as a means of connecting probability-1 assignments to elements of reality, albeit within the limited scope of the interaction events.

\tit{QBism:} When quantum theory is regarded as a decision theory, the fact that it compels an agent to always be uncertain about the outcomes of some measurements cannot directly be taken as a statement characterizing the quantum nature of reality. This is because decision theory presupposes that agents are beings who are perpetually uncertain about things -- if they knew perfectly what would happen as a result of their actions, there would hardly be any use for decision theory! 

Consequently, QBism sees no fundamental difference between the type of uncertainty that occurs in high-precision measurements on quantum systems in the lab, versus the familiar uncertainty of everyday life; both are instances of the same genuine novelty, the world's inexhaustible capacity to surprise us. For the QBist, rather, the difference lies in the particular `shape' of the uncertainty, which is captured by a special equation (the \tit{Urgleichung} \cite{QB_FUCHS_SCHACK}) that becomes relevant for high-precision measurements in the quantum regime. 

Thus while RQM treats `fundamental uncertainty' as a feature that distinguishes quantum systems from classical ones, QBism views uncertainty as referring to an agent's uncertainty about the world, and locates the difference between quantum and classical matter in the particular mathematical tools that agents can use to cope with the uncertainty in each of the two domains.

Finally, since probability assignments are subjective in QBism, probability-1 assignments do not guarantee that the predicted outcome will happen. Thus, unlike in RQM, QBists reject the idea that an agent's probability-1 assignments have any `ontic hold upon the world', and hence they reject the claim that the eigenvalue-eigenstate link has any ontological significance.

This concludes the comparison on the basis of the principles (i)-(iv) of Copenhagenish interpretations. The remaining two sections deal with two further key differences between QBism and RQM.

\section{Is quantum theory law-like or normative? \label{sec:lawsnorms}}

One of QBism's most striking assertions is that the rules of quantum theory do not have the same status as the rules of our other most successful physical theories like classical mechanics or electrodynamics. Whereas those theories give us universal \tit{natural laws} that are supposed to govern the behaviour of all matter, QBism argues that the laws of quantum theory are \tit{normative} rules: an agent measuring what they believe to be quantum systems would be incoherent not to assign probabilities in accordance with quantum theory, but agents may choose to ignore these rules anyway. As we will see, this interpretation of quantum theory puts QBism at odds with RQM, which does not see the laws of quantum theory as being optional or specific to decision-making agents, but rather sees them as universal natural laws governing the production and interactions of physical \tit{information}. In this section we explore the difference between these two viewpoints.

Natural laws, unlike human laws that govern society, are normally conceived by physicists as rules that necessarily govern all material things, human and otherwise. As such, a natural law is meant to be a fact of the universe that all things are forced to obey simply by virtue of being made of matter. Although the process of discovery of natural laws is of course a human endeavour, such that our theoretical models have to be constantly revised and improved, it is nevertheless commonly supposed that scientific progress brings our theories ever closer to some set of natural laws that are ultimately true. Examples of natural laws in physics include Newton's laws of motion and Maxwell's laws of electrodynamics.

By contrast to natural laws, a normative rule is one that does not have to be obeyed. It derives its rule-like character from the idea that it represents the best course of action to take in order to achieve some end. It is conditional in the sense that \tit{if} a decision-making agent wants to achieve some particular end, then they \tit{ought to} obey the rule (but they don't have to). As such, normative rules are only applicable to beings like agents who are directed towards certain ends and capable of choosing their course of action\footnote{QBism does not make any special assumptions about what kinds of things can satisfy these requirements: perhaps machines and other complex arrangements of inanimate matter could qualify. Rather, the question of agency is taken as a pragmatic assumption: so long as it is useful to think of a given entity as choosing actions in order to attain certain ends, then normative rules may be employed in modeling it.}.

It is important to appreciate that in order to say that a theory has a `normative' status, it is not enough merely to point out that the theory is a tool that physicists use to make predictions. This is true of all physical theories, and has no bearing on whether their rules are to be regarded as normative or not. The issue has to do with how the theory explains why its theoretical objects (those things referred to within the theory itself) obey the rules that it posits. 

If the objects of the theory are just physical systems which follow the rules by necessity, then its laws are natural laws. If, on the other hand, the theory explicitly defines decision-making agents as subject to certain imperatives (for instance, self-preservation or rational coherence) and supplies rules that the agents should follow in order to best meet these imperatives, then those rules are normative.

QBism's parting of ways from RQM on this issue can be traced to the central role it gives to agents and the subjective Bayesian interpretation of probability, whereby probabilities quantify an agent's \tit{degrees of belief}. It is easy to appreciate how this leads naturally to the conclusion that quantum theory is normative: it means that the theory is nothing but a set of rules governing how an agent should assign their beliefs about quantum systems\footnote{Note that it is possible to interpret probabilities as objective and still conclude that quantum theory is normative. For instance, in Healey's interpretation probabilities are objective in the sense that there is an objectively `best' quantum state assignment for a given agent. Nevertheless, like in QBism, probabilities are taken to be the agent's degrees of belief, so the rules of quantum theory are normative in Healey's account.}. In fact it can be shown that, given a minimal interpretation of what it means for an agent to believe that a system is quantum\footnote{Formally, this means the agent holds several `generic' beliefs about how systems respond to measurements, as well as the `quantum-specific belief' that the number of outcomes of a minimal informationally complete measurement of the system is equal to the square of it's maximum information storage capacity (i.e.\ its dimension). See \cite{QB_COHERENCE} for details.}, they should allocate their probabilities according to the quantum mechanical Born rule, or else be incoherent~\cite{QB_COHERENCE}. As such, the rules of quantum theory are normative: a rational agent ought to follow them if they want to deal successfully with quantum systems, but they need not do so.

The status of quantum theory in RQM is, at first sight, neither obviously law-like nor obviously normative. According to RQM, quantum theory is not a theory of physical objects or fields evolving in time in the same way as in classical mechanics. Rather, RQM proposes that quantum theory is concerned with processes of \tit{information exchange} between physical systems, which are governed by the abstract rules of information theory and probability theory. One of RQM's goals is to show how the quantum formalism can be derived in this framework from basic postulates of an information-theoretic character \cite{ROVELLI_96,HOEHN1,HOEHN2}. 

However, a closer reading of the literature reveals that RQM is strongly committed to a strictly physical interpretation of `information' (hence probabilities), which in turn is suggestive of taking quantum theory as law-like. For instance, in his earliest paper on RQM Rovelli writes:

\begin{quote}
I shall use here a notion of information that does not require distinction between human and nonhuman observers, systems that understand meaning or do not, very complicated or simple systems, and so on. [\dots] Information expresses the fact that a system is in a certain configuration, which is correlated to the configuration of another system (information source). --- Rovelli, \cite{ROVELLI_96} (p.1653-54).  
\end{quote}

and in a more recent paper, he explains:

\begin{quote}
    `Realism' [in the weak sense] is the assumption that there is a world outside our mind, which exists independently from our perceptions, beliefs or thoughts. Relational QM is compatible with realism in this weak sense. [\dots] There is nothing similar to ‘mind’ required to make sense of the theory. What is meant by a variable taking value ``with respect to a system $S$" is not $S$ to be a conscious subject of perceptions -- it just the same as when we say that the velocity of the Earth is 40km/s ``with respect to the sun": no implication of the sun being a sentient being ``perceiving" the Earth.
\end{quote}

We therefore find that although the rules of quantum theory have an information-theoretic character in RQM, all physical systems must necessarily obey these rules, whether they would like to or not. For instance, in RQM a system can be said to `measure' another system whenever any kind of physical interaction leads to an exchange of information between them; as such, there is no requirement that the measurement be deliberately chosen for any purpose, or even that the system performing it should be aware of performing it. 

We conclude that on RQM's account, quantum theory amounts to a theory of the natural laws of information exchange obeyed by all physical systems. This is in stark contrast to QBism's position, which holds that quantum theory comprises normative rules applicable only to the agents who use the theory, and which they can choose to obey in order to be rationally coherent.

\tit{Remark:} The fact that RQM takes quantum theory as a theory of natural laws is obscured in the literature by a tendency to use language with epistemic overtones. For instance, Rovelli frequently talks about observers having `knowledge', making `predictions' or `asking questions' of another system, even though it is clear that such observers are supposed to include inanimate matter. We take it that such expressions are figures of speech not meant to imply that physical systems can literally make predictions or ask questions in the normal sense; these terms presumably only refer to physical interactions.

\section{Does a shared reality require agreement? \label{sec:agreement}}

In classical mechanics, one and the same event (say, the detection of two test-particles coincident in a small region of space-time) can be described in different ways by different observers. Nevertheless, the event itself is not a \tit{relative fact}: given that it exists relative to any single observer, it must exist in principle relative to all observers. It follows that any difference between observers' descriptions of the event must be traceable to differences between the observers' own standpoints in relation to the event, for instance their usage of different co-ordinate systems. In this manner, it is possible in classical mechanics to interpret every event as an \tit{absolute fact}: stripped of any particular co-ordinates, it stands as a unique element of reality that exists for all observers.

By endorsing the principle of relative facts, QBism and RQM reject the claim that quantum measurement outcomes are absolute facts in the above sense. That is, in at least some circumstances, measurement outcomes must correspond to entirely `private' elements of reality for the relevant observer. As a prototypical example, when Wigner's friend measures the spin of an atom while sealed inside an isolated laboratory, the event corresponding to the outcome of her spin measurement exists only relative to her, so long as she remains sealed in the laboratory.

On the other hand, neither QBism nor RQM goes so far as to say that \tit{all} facts are private: both interpretations allow that facts \tit{can} be shared between observers. For example, Fuchs writes (our emphasis):

\begin{quote}
    The only glaringly mutual world there is for Wigner and his friend in a QBist analysis is the partial one that might come about if these two bodies were to later take actions upon each other (``interact") [\dots] What we learn from Wigner and his friend is that we all have truly private worlds \tit{in addition to our public worlds}. --- \cite{QB_FUCHS_2011}
\end{quote}

How does this `public world' come about? Fuchs, Mermin \& Schack clarify:

\begin{quote}
     An agent-dependent reality is constrained by the fact that different agents can communicate their experience to each other [\dots]. Bob’s verbal representation of his own experience can enter Alice’s, and vice-versa. In this way a common body of reality can be constructed, limited only by the inability of language to represent the full flavor [\dots] of personal experience. --- \cite{QB_FDR2014}
\end{quote}

Addressing the same point in the context of RQM, Rovelli writes:

\begin{quote}
    Does [relative facts] mean that there is no relation whatsoever between views of different observers? Certainly not; [\dots] It \tit{is} possible to compare different views, but the process of comparison is always a physical interaction, and all physical interactions are quantum mechanical in nature. [\dots] Suppose a physical quantity $q$ has value with respect to you, as well as with respect to me. Can we compare these values? Yes we can, by communicating among us. But communication is a physical interaction and therefore is quantum mechanical. --- \cite{ROVELLI_96}
\end{quote}

From these quotes we see that both QBism and RQM are committed to the idea that observers inhabit a shared reality, which is manifested through communication between them; when the sealed laboratory is opened, the friend can then communicate her result to Wigner. But QBism and RQM differ significantly on two important points: first, on the meaning of \tit{communication} between observers; second, on the necessity (or not) of some form of \tit{agreement} in establishing a shared reality.

Regarding the first point, `communication' means something different in QBism than in RQM. For QBists, communication is a means by which agents can make each other aware of their own experiences. In particular, it is a fundamentally normative activity: agents perform certain actions (like vocalizing words belonging to a common language) in order to achieve the goal of reaching mutual understanding, that is, of each agent being able to account for the others' divergent point of view and thereby achieve collective coherence. By contrast, `communication' in RQM is fundamentally a physical interaction between two systems, in which it is irrelevant whether the communication has any social or linguistic meanings attached to it.

This has bearing on the second point, which has to do with whether two agents must necessarily \tit{agree} about any piece of reality to which they share access. For the sake of concreteness, suppose that two agents/observers simultaneously perform measurements on the same system of some physical quantity (eg.\ the spin of an atom in a given direction)\footnote{Note that in order to pose the scenario in QBism we must assume that the agents initially agree that they are measuring the same atom using the same practical procedure. There is no conflict here: nothing in QBism \tit{forbids} agreement; our question is whether there is anything that makes it \tit{mandatory}.}. The question is: do the two parties to the event necessarily agree on the outcome?

Since QBism views communication as a normative activity, it follows that the agents must always be free to disagree. This freedom is important in order to allow the agents to decide which details of their personal experiences are relevant in defining the outcome. For in QBism, a measurement outcome is not necessarily \tit{merely} a bare statement like `the spin is up', but can include additional contextual details referring to each agent's particular experience of the outcome. These might be superficial (eg.\ the agents use different symbols to represent the direction `up') or more profound (eg.\ the agents interpret the measurement procedure using different theoretical models, which leads them to reach different conclusions about what quantity is being measured). Such contextual differences would be manifest in the agents' assigning different POVMs and outcome labels to the event. Hence both agents may disagree on what the measurement outcome was, even while agreeing that it refers to a shared reality, i.e.\ because each one can explain the other's outcome by taking into account their different standpoint and beliefs. Successful communication therefore does not mandate agreement in QBism\footnote{It would be interesting to investigate whether holding certain beliefs might compel agents to agree in order to avoid being incoherent; but that is a different question than the one we are considering here.}.

By contrast, RQM's view of communication as a physical process entails that the observers must `agree' in a stronger sense than in QBism. To see why, it is useful to review in detail how RQM describes the process of an observer $O_1$ measuring a quantity $A$ of system $S$, as defined relative to a second observer $O_2$. 

First, a correlation is established between the values of $A$ and the values of a `pointer variable' of $O_1$, which we denote $V_1$. For concreteness, we can represent the correlation as a bijective map
\eqn{
\phi_1:\trm{dom}(A) \leftrightarrow \trm{dom}(V_1) \, ,
}
which specifies that $O_1$'s pointer displays $V_1 = \phi(a)$ whenever the measured quantity takes value $A=a$. Next, $O_2$ measures the quantity $A$ and obtains $A=a'$. Finally, $O_2$ `communicates' with $O_1$ by measuring $O_1$'s pointer and obtains (say) $V_1 = v''$. From this $O_2$ can infer that the value of $A$ that $O_1$ measured must have been \eqn{
A &=& \phi^{-1}(v'')  \nonumber \\ 
&:=& a''   \, .
}
The issue of `agreement' in RQM then boils down to whether the value which $O_1$ observed (as inferred from their pointer reading $v''$) is \tit{consistent} with the value $a'$ that $O_2$ observed by measuring $A$ directly. It is easy to see that this is the case if and only if the two observed values of $A$ are the same, i.e.\ iff $a'' = a'$. 

Summarizing, if two observers perform the same measurement on the same system and then communicate their results to each other (i.e.\ by interacting with each other's pointer variable), then each observer must necessarily find that both obtained the same measurement outcome. Hence in RQM we can say that any single real event implies a single measurement outcome for all observers who measure it, as judged relative to any one of those observers. In contrast, as we have seen, QBism allows different agents to assign different measurement outcomes to one and the same real event.

\section{Conclusions}

We have seen that QBism and RQM share many features in common by virtue of both being examples of `Copenhagenish interpretations'. Despite these similarities, we have seen that there exist striking incompatibilities between QBism and RQM that go far deeper than just their different preferred interpretations of probability. Taken together, these differences indicate that in spite of their close alignment on many issues, QBism and RQM are not likely to be reconcilable at the level of their underlying ontological commitments. An interesting topic for further research would be to exhibit these differences within an explicit ontological modeling framework capable of encompassing all `Copenhagenish' interpretations.

\acknowledgements

I am grateful to Carlo Rovelli and Andrea Di Biagio for enlightening and cordial correspondence about the differences between RQM and QBism. I also thank Chris Fuchs, Blake Stacey and John DeBrota for feedback on an early draft. This work was supported in part by the John E. Fetzer Memorial Trust.


\end{document}